\documentclass[twocolumn,secnumarabic, showpacs, preprintnumbers, amssymb, floatfix, prb]{revtex4}
\usepackage{epsf}

\begin{document}

\author{I. A. Kokurin}
\email{kokurinia@math.mrsu.ru}
\author{V. A. Margulis}

\affiliation{Institute of Physics and Chemistry, Mordovian State
University, 430000 Saransk, Russia}.

\title{Effect of short-range impurities on low-temperature conductance and
thermopower of quantum wires}

\date{\today}

\begin{abstract}
The electron transport through the parabolic quantum wire placed
in longitudinal magnetic field in the presence of the system of
short-range impurities inside the wire is investigated. Using
approach based on the zero-range potential theory we obtained an
exact formula for the transmission coefficient of the electron
through the wire that allows to calculate such the transport
characteristics as the conductance and differential thermopower.
The dependencies of conductance and thermopower on the chemical
potential and magnetic field are investigated. The effect of
elastic scattering due to short-range impurities on
low-temperature conductance and thermopower is studied. It was
shown that the character of the electron transport essentially
depends on the position of the every scattering center. The
presence even isolated impurity leads to destruction of
conductance quantization. In some cases it is possible that
thermopower can change the sign in dependence on chemical
potential and magnetic field.
\end{abstract}


\maketitle

\section{Introduction}

Electron transport in mesoscopic systems has many long years
attracted attention, both theoretical and experimental. There are
many interesting phenomena in this area. The effect of conductance
quantization has first observed in the narrow constriction that
connect two large areas of two-dimensional electron gas (2DEG) in
high-mobility GaAs-Al$_x$Ga$_{1-x}$As heterostructures
\cite{Wees1988, Wharam1988}. The above discovery has stimulated a
whole series of theoretical works where the various models of
confinment potential have been used: an infinitely long waveguide
with constant cross section \cite{Geyler2000,Bagwell1990},
saddle-point potential \cite{Fertig1987,Buttiker1990} and other
one \cite{Torres1994,Bogachek1994,Maao1994}. All these papers
based on Landauer-B\"{u}ttiker transport approach
\cite{Landauer1987,Buttiker1985}.

The character of conductance quantization in quasi-one-dimensional
(Q1D) systems depends on the geometry of the quantum wire, on its
length and width \cite{Glazman1988}. Applied magnetic field {\bf
B} plays the key role in a such system. A magnetic field enhances
lateral confinement so that by varying {\bf B} we can alter the
effective geometrical size of the system, and hence the functional
dependence of the conductance on the field {\bf B}. In particular,
by varying {\bf B} one can change the parameters of the
conductance quantization steps (length of the conductance plateau)
\cite{Wees1991}. The parameters of the conductance quantization
depend sufficiently not only on magnitude of magnetic field but
also on its direction \cite{Geyler2000}.

However, the conductance quantization is very sensitive to
electron scattering. There are three principal type of scattering
that can lead to destruction of the conductance quantization: (i)
scattering in the region of the contact of wire with electron
reservoirs \cite{Kozlov2007,Garcia1989}, (ii) electron scattering
on rough walls of the wire \cite{Leng1993} and (iii) impurity
scattering inside the Q1D wire (see, for instance,
Ref.\onlinecite{Sukhorukov1994} and references therein).
Interference of electron waves due to every above-mentioned types
of scattering leads to oscillatory dependence of conductance on
Fermi energy.

Simultaneous consideration of every above-mentioned scattering
factors is the sufficiently complicated problem. Therefore in the
present paper, we shall focus our attention only on the effect of
impurity scattering.

Note that all previous papers concerned consideration of the
effects of the impurities on the electron transport in Q1D systems
had some limitations: as a rule the number of impurities was equal
to unity. Both experimental \cite{Faist1990} and theoretical
\cite{Bagwell1990,Bagwell1992, Levinson1991} study of the effect
of impurity on quantized conductance of Q1D wire shows that even
isolated impurity can lead to fundamental change of the conduction
character.

Large number of works were devoted to conductance of quantum wires
in the case when an isolated short-range impurity contained inside
the wire \cite{Levinson1991, Chu1989}. It was found that the main
features of the conductance $G$ are downward dips in the graph of
$G$ vs Fermi energy below the bottom of each transverse mode in
the case of infinite rectilinear waveguide \cite{Chu1989} and a
resonance peak below the first quantized step in the case of a
saddle-point potential of quantum point contact (QPC) due to
resonant transmission through bound state \cite{Levinson1991}. The
effect of a finite-range scatterer was discussed in Refs.
\onlinecite{Bagwell1992,Kim1999}.

Besides conductance $G$, thermopower $S$ is another transport
property of Q1D systems in which quantization effect manifests. It
was first shown theoretically by Streda \cite{Streda1989} that in
narrow constriction the thermopower $S$ exhibits oscillatory
dependence as a function of Fermi energy. It was shown that peak
values of $S$ are quantized, given by $\ln 2/(i+1/2)$ (in units
$k_B/e$). These peaks occur when the Fermi energy $\mu$ is close
to edge of $i$th transverse subband ($i=1,2,...$), that is the
peak in thermopower corresponds to the quantized conductance
threshold.

The above-mentioned quantum oscillations have been demonstrated in
experiment \cite{Molenkamp1990,Appleyard1998,Seol2007}. Later
theoretical studies \cite{Proetto1991,Margulis2003,Kokurin2004}
confirm the results of Streda and moreover dependencies of
thermopower on magnitude and direction of applied magnetic field
were studied \cite{Kokurin2004}.

However we know only one paper where effect of impurity on
low-temperature thermopower was investigated \cite{Chu1994}. The
existence of negative thermopower due to backscattering was shown.

We present the theory of the quasi-ballistic electron transport in
quantum wire for the case of any finite number of scattering
centers. The aim of present work is to study the effect of elastic
scattering due to $N$ short-range impurities on the such transport
characteristics as conductance and thermopower of quantum wire. We
include in consideration only effects those deal with impurity
scattering excluding other scattering types.

The paper is organized as follows. In Sec. 2 we describe the model
and Hamiltonian and find analytical formula for transmission
amplitudes. In Sec. 3 we remind reader the formulas for the
conductance and thermopower and show character features of
impurity-assisted conductance and thermopower. Sec. 4 consists of
brief summary.

\section{Hamiltonian and transmission coefficients}

Now we consider the uniform quantum wire with constant
cross-section in presence of homogeneous longitudinal magnetic
field ${\bf B}=(0,0,B)$. Unperturbed one-electron spinless
Hamiltonian in quantum wire given by
\begin{equation}
\label{H_0}
H_0=\frac{1}{2m^*}\left(\mathbf{p}-\frac{e}{c}\mathbf{A}\right)^2+V(x,y),
\end{equation}
where $m^*$ is the electron effective mass, ${\bf A}$ is the
vector-potential of the field ${\bf B}$, $V(x,y)$ is the
confinment potential in $(x,y)$ -- plane. In our model the
confinment potential $V(x,y)$ is symmetric parabolic potential
$V(x,y)=m^*\omega_0^2(x^2+y^2)/2$, where $\omega_0$ is the
characteristic frequency of the parabolic potential that
determines the effective radius of the wire
$r_0=\sqrt{\hbar/m^*\omega_0}$. It is convenient to chose the
vector-potential in symmetric gauge ${\bf A}=(-yB/2,xB/2,0)$.

The spectrum of the Hamiltonian (\ref{H_0}) with the symmetric
parabolic potential $V(\rho)=m^*\omega_0^2\rho^2/2$ has the
well-known form
\begin{equation}
\label{spectrum}
 E_{mnk}={\hbar\omega_c\over2}m+
{\hbar\Omega\over2}(2n+|m|+1)+\frac{\hbar^2k^2}{2m^*},
\end{equation}
where $\Omega=\sqrt {\omega_c^2 +4\omega_0^2}$,
$\omega_c=|e|B/m^*c$ is the cyclotron frequency, $m=0,\pm 1,\pm
2,...$, $n=0,1,2,...$, $k$ is the electron wavevector in wire's
axis direction.

The corresponding wavefunctions in cylindrical coordinates ${\bf
r}=(\rho, \varphi, z)$ are given by
\begin{equation}
\label{psi_0}
\psi_{mnk}^0=R_{mn}(\rho){e^{im\varphi}\over\sqrt{2\pi}}e^{ikz},
\end{equation}
with
\[
R_{mn}(\rho)=C_{mn}\rho^{|m|}\exp(-\rho^2/4l_0^2)L_n^{|m|}(\rho^2
/{2l_0^2)},
\]
and
\[ C_{mn}={1\over{l_0^{|m|+1}}}\left[\frac{n!}
{2^{|m|}(n+|m|)!}\right]^{1/2},
\]
where $L_n^{|m|}(x)$ are generalized Laguerre polynomials,
$l_0=\sqrt{\hbar/ m^*\Omega}$. It is evident the normalization of
the longitudinal part of the wavefunction (\ref{psi_0}) can be
chosen in arbitrary form, because it is unimportant to find
transmission coefficients.

It is well-known that the Green function $G({\bf r}, {\bf r}'; E)$
can be represented for Q1D system in the following form
\begin{equation}
G({\bf r}, {\bf r}'; E)= \sum\limits_\alpha G_z(z,z';E-E_\alpha)
\psi_\alpha(x,y)\psi_\alpha^*(x',y'),
\end{equation}
where $\psi_\alpha(x,y)$ is the transverse part of Q1D
wavefunction, $E_\alpha$ is the bottom of transverse subband
$\alpha$, and
$$
G_z(z, z'; E)={i m^*\over\hbar^2 k}e^{ik|z-z'|}
$$
is the Green function of the operator
$H_z=-(\hbar^2/2m^*)\partial^2/\partial z^2$.

In the case of our model above-mentioned representation leads to
the following form of Green's function of the unperturbed
Hamiltonian (\ref{H_0}), that is necessary to solve the present
problem in some analogy with Ref.\onlinecite{Levinson1991}
\begin{eqnarray}
\label{Green}
\nonumber
G({\bf r}, {\bf r}'; E)&=&\frac{i m^*}{2
\pi \hbar^2}\sum\limits_{n=0}^{\infty}
\sum\limits_{m=-\infty}^{\infty}R_{mn}(\rho)R_{mn}(\rho')\\
&&\times\frac{e^{ik_{mn}|z-z'|+im(\varphi-\varphi')}}{k_{mn}},
\end{eqnarray}
where $\hbar k_{mn}=\sqrt{2m^*(E-E_{mn})}$.

Let us find the exact solution of Schr\"{o}dinger equation for the
Hamiltonian $H=H_0+U({\bf r})$, where $U({\bf r})$ is the
potential created by $N$ short-range impurities. It is rightful to
use the zero-range potential \cite{Demkov1975} for modelling
short-range impurities as long as the impurity range is shorter
than all other relevant length scales in the problem. Let us
remind that above potential describes only spherically-symmetric
$s$-scatterer and can be expressed as the boundary condition for
the wavefunction \cite{Demkov1975}.

This approach allows to find exact solution of Schr\"{o}dinger
equation and therefore we can find the exact expressions for
transmisssion coeficients that in one's turn allows to find
conductance and thermopower \cite{Sivan1986} (see Eqs.(\ref{G})
and (\ref{S})).

The exact solution of Schr\"{o}dinger equation with Hamiltonian
$H=H_0+U({\bf r})$ can be found with the help of technique based
on the theory of self-adjoint extensions of symmetric operators
(see for instance Ref. \onlinecite{Bruning2003}). Here we use only
some consequences of above-mentioned approach. The asymptotic of
the wave-function in the vicinity of the each impurity point ${\bf
q}_j$ does not depend on any smooth potential \cite{Demkov1975}
(in our case it is the confinment potential) and has the form
\begin{equation}
\psi({\bf r})|_{{\bf r}\rightarrow {\bf q}_j}=C_j\left (
\frac{1}{|{\bf r}-{\bf q}_j|}-\frac{1}{a_j}\right)+O(|{\bf r}-{\bf
q}_j|),
\end{equation}
where $C_j$ is the some constant, $a_j$ is the scattering length
for impurity that locates at the point ${\bf q}_j$.

The zero-range potential theory allows us to represent the
solution of Schr\"{o}dinger equation with Hamiltonian $H$ in terms
of the Green's function $G({\bf q}_i,{\bf q}_j;E)$
\cite{Bruning2003}
\begin{equation}
\label{psi_alpha} \psi({\bf r})=\psi^0({\bf
r})+\sum_{j=1}^N\alpha_jG({\bf r},{\bf q}_j;E),
\end{equation}

Considering the asymptotic of the eigenfunction (\ref{psi_alpha})
near each point ${\bf q}_j$ one can find the system equation for
$\alpha_j$ determination
\begin{equation}
\label{Eq_sys} \sum_{i=1}^N\alpha_i({\cal
Q}_{ij}+\Lambda_j\delta_{ij})+\psi^0({\bf
q}_j)=0,\;\;\;\;\;j=1,...,N.
\end{equation}
where $\Lambda_j$ is the parameter that characterizes the strength
of the zero-range potential at the point ${\mathbf q}_j$ and that,
with respect to the scattering length $a_j$, is given by equation
$\Lambda_j=m^*/2\pi\hbar^2a_j$. Negative $\Lambda$ corresponds to
presence of the bound state for isolated impurity (whereas for the
case $N>1$ bound states can exist for either sign of $\Lambda$).
Large values of $|\Lambda|$ correspond to weak impurity. The
elements of so-called Krein's $\cal{Q}$-matrix are given by
\[
{\cal Q}_{ij}(E)=\cases{G({\bf q}_i,{\bf
q}_j;E),\;\;\;\;\;\;\;\;\;\;\;\;\;\;\;\;\;\;\;\;\;\;\;\;\;\;\;\;\;\;\;\;\
\ i\neq j, \cr
 \lim\limits_{{\bf r}\rightarrow {\bf q}_j}\left[G({\bf r},{\bf
q}_j;E)-\frac{m^*}{2\pi\hbar^2}\frac{1}{|{\bf r}-{\bf q}_j|}
\right],\ \ \ i=j.  \cr}
\]

The convergence of ${\cal Q}_{jj}$ was demonstrated in
Ref.\onlinecite{Geyler1997} by means the integral representation.

The asymptotic expansion for $G({\bf r},{\bf q}_j;E)$ at ${\bf
r}\rightarrow {\bf q}_j$ can be found only in some limiting cases.
However the impurity scattering is more sufficient when the
impurities are located in the vicinity of wire's axis
\cite{Geyler1997}. In this case we can use the following formula
for ${\cal Q}_{jj}(E)$ determination
\begin{equation}
{\cal Q}_{jj}(E)= \lim\limits_{{\bf r}\rightarrow {\bf
q}_j}\left[G({\bf r},{\bf q}_j;E)-G({\bf r},{\bf
q}_j;E_0)\right]+{\cal C},
\end{equation}
where $E_0$ is the some fixed value of the energy, the constant
${\cal C}$ can be determined considering case when the asymptotic
of Green's function is definite. It is conveniently to put
$E_0=0$. In this case ${\cal C}$ is given by ${\cal C}=(m^*/2\sqrt
2\pi\hbar^2l_0)\zeta(1/2;1/2)$, where $\zeta(s;x)$ is the
generalized Riemann zeta function.

The solution of  system of equations (\ref{Eq_sys}) leads us to
the eigenfunctions of the Hamiltonian $H$ (with the same energy
that for $\psi^0_{mnk}$)
\begin{equation}
\label{psi} \psi({\bf r})=\psi^0({\bf r})-\sum_{i,j=1}^N[{\cal
Q}(E)+{\cal I}\Lambda]^{-1}_{ij}\psi^0({\bf q}_j)G({\bf r},{\bf
q}_i;E).
\end{equation}
Here ${\cal I}$ is the unitary matrix $N\times N$,
$\Lambda=(\Lambda_1,...,\Lambda_N)^T$ ('T' denotes the transposed
vector), ${\cal A}^{-1}$ is the inverse matrix for ${\cal A}$.

It is evident that far from impurities ($z\rightarrow\infty$) the
following expansion is right
\begin{equation}
\label{expansion}
\psi_{mnk}({\bf r})=\sum_{m'n'} t_{mn,m'n'}
\psi^0_{m'n'k'}({\bf r}),
\end{equation}
where $t_{mn,m'n'}$ is the transmission amplitude from incident
mode $mn$ to transmitted mode $m'n'$ ($k'=k_{m'n'}$), and the
summation in Eq.(\ref{expansion}) is over all transmitted modes
($E>E_{mn}$, for real $k_{mn}$) because the modes which correspond
to imaginary $k_{mn}$ rapidly decay on large $z$. From
Eqs.(\ref{psi},\ref{expansion}) we find for the transmission
amplitude
\begin{eqnarray}
\label{tr_amplitude}
\nonumber
&& t_{mn,m'n'}=\delta_{mm'}\delta_{nn'}-\frac{im^*}{2\pi\hbar^2k_{m'n'}}\\
&&\times\sum_{i,j=1}^N\left(\psi^0_{m'n'k'}({\bf
q}_i)\right)^*[{\cal Q}(E)+{\cal I}
\Lambda]^{-1}_{ij}\psi^0_{mnk}({\bf q}_j).
\end{eqnarray}

It is conveniently to write Eq.(\ref{tr_amplitude}) in the matrix
form
\begin{equation}
t_{mn,m'n'}=\delta_{mm'}\delta_{nn'}-\frac{im^*}{2\pi\hbar^2k_{m'n'}}
\widehat{\psi}_1\widehat{F}\widehat{\psi}_2,
\end{equation}
where $\widehat{\psi}_1=(\psi^0_{m'n'k'}({\bf
q}_1),...,\psi^0_{m'n'k'}({\bf q}_N))$, $\widehat{F}$ is the
inverse matrix for $[\widehat{{\cal Q}}(E)+\widehat{{\cal I}}
\widehat{\Lambda}]$, and $\widehat{\psi}_2=(\psi^0_{mnk}({\bf
q}_1),...,\psi^0_{mnk}({\bf q}_N))^T$.

The first term in Eq.(\ref{tr_amplitude}) corresponds to ballistic
transmission of electrons through the wire \cite{Geyler2000} when
the inter-subband transitions are forbidden. The second term
corresponds to mode-mixing due to impurity scattering.

\section{Conductance and Thermopower}

Accordingly to Landauer-B\"{u}ttiker approach the two-probe
Landauer conductance $G$ is given by
\cite{Buttiker1985,Landauer1987}
\begin{equation}
\label{G} G=\frac{e^2}{\pi\hbar}\int_0^\infty
dE\left(-\frac{\partial f}{\partial E}\right)\sum_{\alpha,\alpha'}
T_{\alpha,\alpha'}(E),
\end{equation}
where $f(\varepsilon)=[e^{(\varepsilon-\mu)/T}+1]^{-1}$ is the
Fermi distribution function, $\mu$ is the chemical potential.

For the case of thermoelectric transport the differential
thermopower $S$ is given by the formula \cite{Sivan1986}
\begin{equation}
\label{S} S=\frac{k_B}{e}\frac{\int_0^\infty
dE\left(-\frac{\partial f}{\partial
E}\right)\left(\frac{E-\mu}{T}\right)\sum_{\alpha,\alpha'}T_{\alpha,\alpha'}(E)}{\int_0^\infty
dE\left(-\frac{\partial f}{\partial
E}\right)\sum_{\alpha,\alpha'}T_{\alpha,\alpha'}(E)}.
\end{equation}

The transmission coefficients which corresponds to the electron
transition from transverse sub-band $\alpha$ to sub-band $\alpha'$
are
\begin{equation}
\label{T_aa'}
T_{\alpha\alpha'}=\frac{k_{\alpha'}}{k_\alpha}|t_{\alpha\alpha'}|^2.
\end{equation}

For the case of comparatively low temperatures ($T\lesssim 5$K)
the so-called Mott formula is a good approximation for thermopower
\cite{Lunde2005}
\begin{equation}
\label{S^M} S^M=\frac{k_B}{e}\frac{\pi^2T}{3}\frac{\partial \ln
G(\mu,T)}{\partial \mu}.
\end{equation}

For the case of the ultra-low temperatures ($T\lesssim 1K$) there
are more rough approximation. In this case Eq.(\ref{S^M}) include
$G(\mu,T=0)$. One can use this approximation in order to analyze
impurity-assisted features in thermopower because they have a
place at ultra-low temperatures. This approximation allows to
analyze the low-temperature thermopower except the points in
vicinity of conductance threshold where the zero-temperature
conductance has the breakdown of the derivative on chemical
potential.

Combining the Eqs.(\ref{T_aa'}),(\ref{G}) and (\ref{tr_amplitude})
we can calculate the quasi-ballistic conductance of quantum wire.
Let us consider numerically the case of $N$ impurities. From now
on we use for electron effective mass $m^*=0.067m_e$. On the Fig.1
the conductance of quantum wire that contains $N$ impurities
($N=1,2,3$) is plotted. The various curves plotted for various
quantity, strength and spatial distribution of impurities. One can
see from Fig.1 that conductance crucially depend on the position
of every scattering center, its quantity and strength.

\begin{figure}[!h]
{\centering
\epsfclipon
\epsfxsize=84mm
\caption{\label{fig1} Zero-temperature conductance as a function
of Fermi energy plotted for various number, strength and spatial
distribution of impurities. Dashed lines correspond to ballistic
conductance of the perfect wire without impurities.
$\omega_0=10^{13}$s$^{-1}$, $B=1$T.}}
\end{figure}

The analysis shows that $G$ equals $n(e^2/\pi\hbar)$ whenever the
Fermi level is at the band bottom of the $(n+1)$th subband,
irrespective of the strength and location of the scatterers. And
there is the zero of upper-mode transmission coefficient when the
Fermi energy tends to $(E_{mn}+0)$ for any strength of the
scatterers. For the purpose of comparison, the perfect-wire
results are also plotted near all lines and they are indicated by
dashed lines.

Taking into account that all character features of
impurity-assisted thermopower appear at ultra-low temperature we
can use the Cutler-Mott type formula (where zero-temperature
conductance) everywhere except for the thresholds of conductance
quantization where conductance-derivative undergoes the breakdown.

\begin{figure}[!h]
{\centering
\epsfclipon
\epsfxsize=84mm
\caption{\label{fig2} Low-temperature thermopower as a function of
Fermi energy plotted for various number, strength and spatial
distribution of impurities. Dashed lines correspond to thermopower
of the perfect wire without impurities. All parameters are the
same as in Fig.~\ref{fig1}.}}
\end{figure}

One can see from Fig.2 that thermopower $S$ as the conductance $G$
crucially depends on the position of every scattering center, its
quantity and strength.

One can see that there is negative $S$ at $\mu$ that corresponds
negative tilt of the curve $G(\mu)$.

It is well-known that even single impurity influences hardly on
the carrier transport through one-dimensional system and it can
even lead to destruction of conductance quantization. For the case
of single impurity the Krein's ${\cal Q}$-matrix is the scalar
function and therefore we can find the exact formula for
conductance at temperature $T=0$
\begin{equation}
\label{G_1imp} \frac{G(\mu,T=0)}{G_0}={\cal N}(\mu)-\frac{{\rm
Im}^2{\cal Q}(\mu)}{|{\cal Q}(\mu)+\Lambda|^2}.
\end{equation}
Here ${\cal N}(\mu)$ is the number of transverse-quantization
sub-band under the Fermi energy and $G_0=e^2/\pi\hbar$ is the
conductance quantum. In Eq.(\ref{G_1imp}) ${\cal N}(\mu)$
describes the step-like dependence of ballistic conductance at
$T=0$.

For the case of one impurity we have downward dips near the
thresholds in the graph of $G$ vs Fermi energy (see Fig.1) (if
bound state center exist, $\Lambda<0$). In opposite case
($\Lambda>0$) the dependence $G(\mu)$ is monotonic but it is not
step-like.

The energies for which ${\rm Im}Q(\mu)=0$ corresponds full
transmission. And case when ${\rm Re}Q(\mu)+\Lambda=0$ corresponds
to the conductance minima $G/G_0={\cal N}(\mu)-1$.

The solution of above equation ${\rm Re}Q(E_0)+\Lambda=0$
corresponds to a first approximation for energy of
quasi-bound-state $E-i\Gamma$ on the complex energy plain, where
$\Gamma$ determines width of the quasi-bound level.

For the case of several impurities ($N>1$) there are the strong
dependence of transmission coefficients on the distance between
each pair of impurity centers $|{\bf q}_i-{\bf q}_j|$
($i,j=1,...,N$). Thus we have the strong dependence of $G$ and $S$
on position of each impurity ${\bf q}_j$.

\section{Summary}

We present the exact solution for the problem of the transmission
of charge-carrier through the quantum wire that contain the system
of short-range impurities. Accordingly the Landauer--B\"{u}ttiker
approach conductance and thermopower of above system are
investigated. It is obviously that in general, conductance $G$ is
lowered due to impurity scattering. The interference phenomena in
transmission and existence of the negative thermopower were shown.

Note that in the case of two impurities the distance $|{\bf
q}_1-{\bf q}_2|$ is equal to wire's length one can speak about
modelling of the weak scattering in the region of wire-reservoir
contacts.

Let us note that for the case constriction model there are not
limitation on the electron spectrum from the bottom
\cite{Levinson1991} in constraint to the case of uniform wire.
That is why the level of bound state on the impurity is in
continuous spectrum and it is possible the resonant transmission.

Note that $s$-scattering can not explain all features of the
scattering in Q1D systems, but on the whole this one is in good
qualitative agreement with experimental data.


\begin{thebibliography}{99}

\bibitem{Wees1988} B. J. van Wees, H. van Houten, C. W. J.
Beenakker, J. G. Williamson, L. P. Kouwenhoven, D. van der Marel,
and C. T. Foxon, Phys. Rev. Lett. {\bf 60}, 848 (1988).

\bibitem{Wharam1988} D. A. Wharam, T. J. Thornton, R. Newbury,
M. Pepper, H. Ahmed, J. E. F. Frost, D. G. Hasko, D. C. Peacock,
D. A. Ritchie, and G. A. C. Jones, J. Phys. C: Solid State Phys.
{\bf 21}, L209 (1988).

\bibitem{Geyler2000} V. A. Geyler, V. A. Margulis, Phys. Rev. B {\bf 61}, 1716 (2000).

\bibitem{Bagwell1990} P. F. Bagwell, Phys. Rev. B {\bf 41}, 10354
(1990).

\bibitem{Fertig1987} H. A. Fertig and B. I. Halperin, Phys. Rev. B {\bf
36}, 7969 (1987).

\bibitem{Buttiker1990} B\"uttiker M, Phys. Rev. B {\bf 40}, 7906 (1990).

\bibitem{Torres1994} J. A. Torres, J. I. Pascual, and J. J. S\'aenz,
Phys. Rev. B {\bf 49}, 16581 (1994).

\bibitem{Maao1994} F. A. Maa\o{}, I. V. Zozulenko, and E. H. Hauge, Phys. Rev. B, {\bf 50},
17320 (1994).

\bibitem{Bogachek1994} E. N. Bogachek, M. Jonson, R. I. Shekhter,
and T. Swahn, Phys. Rev. B {\bf 50}, 18341 (1994).

\bibitem{Landauer1987} R. Landauer, Z. Phys. B {\bf 68}, 217 (1987).

\bibitem{Buttiker1985} M. B\"uttiker, Y. Imry, R. Landauer, and S. Pinhas, Phys. Rev.
B {\bf 31}, 6207 (1985).

\bibitem{Glazman1988} L. I. Glazman, G. B. Lesovik, D. E.
Khmel'nitskii, and R. I. Shekhter, Pis'ma Zh. Eksp. Teor. Fiz.
{\bf 48}, 218 (1988) [JETP Lett. {\bf 48}, 238 (1988)].

\bibitem{Wees1991} B. J. van Wees, L. P. Kouwenhoven, E. M. M. Willems,
C. J. P. M. Harmans, J. E. Mooij, H. van Houten, C. W. J.
Beenakker, J. G. Williamson, and C. T. Foxon, Phys. Rev. B {\bf
43}, 12431 (1988).

\bibitem{Kozlov2007} D. A. Kozlov, Z. D. Kwon, A. E. Plotnikov, D. V.
Shcheglov, A. V. Latyshev, Pis'ma Zh. Eksp. Teor. Fiz. {\bf 86},
752 (2007) [JETP Lett. {\bf 86}, ??? (2007)].

\bibitem{Garcia1989} N. Garc\'{\i}a and L. Escapa, Appl. Phys. Lett. {\bf
54}, 1418 (1989); L. Escapa and N. Garc\'{\i}a, Appl. Phys. Lett.
{\bf 56}, 901 (1990).

\bibitem{Leng1993} M. Leng, and C. S. Lent, Phys. Rev. Lett. {\bf 71},
137 (1993); Phys. Rev. B {\bf 50}, 10823 (1994); K. Nikoli\'{c},
and A. MacKinnon, Phys. Rev. B {\bf 50}, 11008 (1994); K.
Nikoli\'{c}, and R. \v{S}ordan, Phys. Rev. B {\bf 58}, 9631
(1998).

\bibitem{Sukhorukov1994} E. V. Sukhorukov, M. I. Lubin, C. Kunze, Y. B. Levinson,
Phys. Rev B {\bf 49}, 17191 (1994).

\bibitem{Faist1990} J. Faist, P. Gu\'{e}ret, and H. Rothuizen,
phys. Rev. B {\bf 42}, 3217 (1990); P. L. McEuen, B. W. Alphenaar,
R. G. Wheeler, and R. N. Sacks, Surf. Sci. {\bf 229}, 312 (1990);
D. H. Cobden, N. K. Patel, M. Pepper, D. A. Ritchie, J. E. F.
Frost, and G. A. C. Jones, Phys. Rev. B {\bf 44}, 1938 (1991).

\bibitem{Levinson1991} Y. B. Levinson, M. I. Lubin, and E. V. Sukhorukov,
Pis'ma Zh. Eksp. Teor. Fiz. {\bf 54}, 405 (1992) [JETP Lett. {\bf
54}, 401 (1992)]; M. I. Lubin, Pis'ma Zh. Eksp. Teor. Fiz. {\bf
57}, 346 (1993) [JETP Lett. {\bf 57}, 361 (1993)]; Y. B. Levinson,
M. I. Lubin, E. V. Sukhorukov, Phys. Rev. B {\bf 45}, 11936
(1992).

\bibitem{Bagwell1992} P. F. Bagwell and R. K. Lake, Phys. Rev. B {\bf
46}, 15329 (1992).

\bibitem{Chu1989} C. S. Chu, R. S. Sorbello, Phys. Rev. B {\bf 40},
5941 (1989).

\bibitem{Kim1999} C. S. Kim, and A. M. Satanin, Zh. Eksp. Teor. Fiz. {\bf 115},
211 (1999) [JETP {\bf 88}, 118 (1999)].

\bibitem{Streda1989} P. Streda, J. Phys.: Condens. Matter {\bf 1},
1025 (1989).

\bibitem{Appleyard1998} N. J. Appleyard, J. T. Nickolls, M. Y. Simmons, W. R. Tribe,
and M. Pepper, Phys. Rev. Lett. {\bf 81}, 3491 (1998).

\bibitem{Molenkamp1990} L. W. Molenkamp, H. and van Houten, C. W. J. Beenakker,
R. Eppenga, and C. T. Foxon, Phys. Rev. Lett. {\bf 65}, 1052
(1990).

\bibitem{Seol2007} J. H. Seol, A. L. Moore, S. K. Saha, F. Zhou, Li
Shia, Q. L. Ye, R. Scheffler, N. Mingo, and T. Yamada, J. Appl.
Phys. {\bf 101}, 023706 (2007).

\bibitem{Proetto1991} C. R. Proetto, Phys. Rev. B {\bf 44}, 9096 (1991).

\bibitem{Margulis2003} V. A. Margulis and A. V. Shorokhov, J. Phys.: Condens. Matter {\bf 15},
4181 (2003).

\bibitem{Kokurin2004} I. A. Kokurin, V. A. Margulis, and A. V.
Shorokhov, J. Phys.: Condens. Matter {\bf 16}, 8015 (2004).

\bibitem{Chu1994} C. S. Chu,  M.-H. Chou, Phys. Rev. B {\bf 50},
14212 (1994).

\bibitem{Demkov1975} Yu. N. Demkov and V. N. Ostrovsky,
{\it Zero-Range Potentials and Their Application in Atomic
Physics} (Leningrad, Leningrad State University Press, 1975)
[Plenum, NewYork, 1988].

\bibitem{Sivan1986} U. Sivan and Y. Imry, Phys. Rev. B {\bf 33}, 551
 (1986).

\bibitem{Bruning2003} J. Br\"uning and V. A. Geyler, J. Math. Phys. {\bf 44}, 371 (2003).

\bibitem{Geyler1997} V. A. Geiler and V. A. Margulis, Zh. Eksp. Teor. Fiz. {\bf 111},
2215 (1997) [JETP {\bf 84}, 1209 (1997)].

\bibitem{Lunde2005} A. M. Lunde and K. Flensberg, J. Phys.: Condens. Matter {\bf 17},
3879 (2005).

\end{thebibliography}
\end{document}